\documentclass[apj,twocolumn]{openjournal}

\usepackage{xcolor}
\usepackage{textgreek}
\usepackage[utf8]{inputenc}
\usepackage[english]{babel}
\usepackage{graphicx} 

\usepackage{hyperref}
\hypersetup{
    unicode, 
    colorlinks=true,
    linkcolor=linkcolor,
    citecolor=linkcolor,
    filecolor=linkcolor,
    urlcolor=linkcolor,
}
\usepackage{color,colortbl}
\definecolor{linkcolor}{rgb}{0.0,0.3,0.5}
\usepackage{tensind}
\tensordelimiter{?}
\DeclareGraphicsExtensions{.bmp,.png,.jpg,.pdf}
\usepackage{verbatim}
\usepackage[normalem]{ulem}
\usepackage{orcidlink}
\usepackage{soul}

\graphicspath{ {./figs/} }

\begin{document}

\title{Pulsar Science with the SKA Observatory}

\author{Bhal Chandra Joshi$^{1,2}$\orcidlink{0000-0002-0863-7781}}

\affiliation{$^{1}$National Centre for Radio Astrophysics, SP Pune University Campus, Pune 411007, Maharashtra, India}
\affiliation{$^{2}$Department of Physics, Indian Institute of Technology Roorkee, Roorkee 247667, Uttarakhand, India}

\author{Aris Karastergiou$^{3}$}

\affiliation{$^{3}$Department of Astrophysics, University of Oxford, Denys Wilkinson Building, Keble Road, Oxford OX1 3RH, UK}

\author{Marta Burgay$^{4}$\orcidlink{0000-0002-8265-4344}}

\affiliation{$^4$INAF - Osservatorio Astronomico di Cagliari, via della Scienza 5, 09047 Selargius (CA), Italy}

\author{The SKA pulsar science working group}

\begin{abstract}
    The large instantaneous sensitivity, a wide frequency coverage and flexible observation modes with large number of beams in the sky are the main features of the SKA observatory's two telescopes, the 
    SKA-Low and the SKA-Mid, which are located on two different continents. Owing to these capabilities, the SKAO telescopes are going to be a game-changer for radio astronomy in general and pulsar astronomy in particular. The eleven articles in this special issue on pulsar science with the SKA Observatory  describe its impact on different areas of pulsar science. In this lead article, a brief description of the two telescopes highlighting the relevant features for pulsar science is presented followed by an overview of each accompanying article, exploring the inter-relationship between different pulsar science use cases.  
\end{abstract}

\section{Introduction}

The SKA Observatory (SKAO) is going to be a game-changer 
for radio astronomy in the coming decade. The construction of 
this  unique instrument, comprising of two 
telescopes in two different continents, has already commenced with 
the phase 1 of these telescopes providing an unprecedented seamless 
frequency coverage from 50 MHz to 15 GHz with unmatched sensitivities 
in highly flexible interferometric configurations. Therefore, the SKAO  
is expected to enable a range of diverse astronomical investigations 
with significant impact on fundamental aspects of astrophysics.

\begin{figure*}[t] 
  \centering
  \includegraphics[width=\textwidth]{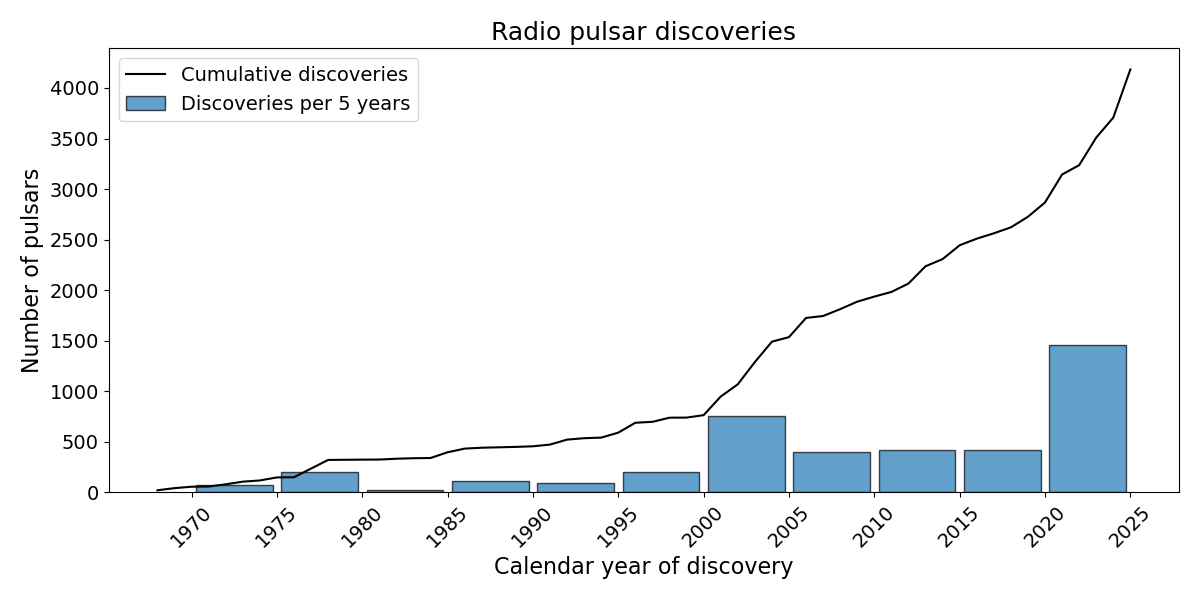}
  \caption{The discovery of pulsars has approximately occurred  
  at an exponential rate with time, matching developments in new 
  telescope hardware and software. The significant increase 
  in the last decade corresponds to commencement of operation 
  of upgraded GMRT, MeerKAT and FAST instruments. Current telescopes and future 
  assemblies of the SKAO will continue this trend.}
  \label{fig:discoveries}
\end{figure*}

One such area of radio astronomy, where the SKAO is expected to make 
fundamental discoveries, is pulsar astronomy. Historically, new instruments 
have accelerated the pace of progress in pulsar astronomy (Figure \ref{fig:discoveries}) 
and the SKAO telescopes are likely to drive significant progress in this 
field in the next decade. The phase 1 of the rollout 
of the SKAO telescope, called AA4 (Array Assembly 4), is likely to double the known pulsar 
population in new surveys \citep{keane2025,abbate2025,bagchi2025}. 
These new discoveries will improve our understanding of the dynamics, 
evolution  and gas content of globular clusters \citep{bagchi2025} and 
the black hole at the centre of the Milky Way galaxy \citep{abbate2025} 
apart from increasing the samples for each of different kinds of 
radio emitting neutron stars (NS), making it possible to uncover 
evolutionary pathways between different types of NS \citep{levin2025}. 
The larger population sample will enhance our 
understanding of the magneto-ionic interstellar medium  
\citep{tiburzi2025,xu2025}, the pulsar magnetosphere 
\citep{oswald2025} and pulsar wind nebulae  \citep{gelfand2025}. 
Moreover, the discovery of exotic neutron star systems  will test 
gravity theory ever more stringently \citep{venkatraman2025}  
and will probe fundamental physics at sub-atomic
level \citep{basu2025}. Finally, this enhanced sample is likely to 
make the sky portrait sharper in nano-Hertz gravitational waves 
impacting on our understanding of the Universe in a fundamental 
way \citep{shannon2025}. In summary, the papers in this special 
issue describe the way the upcoming  SKA Observatory's telescopes 
address fundamental physics through the study of pulsars and gravitational 
waves.


The SKAO has been in planning over the last two decades and 
pulsar astronomy itself has evolved over these decades, often driven 
by new discoveries by the SKA precursor and pathfinder instruments. 
Consequently, the wish list of pulsar astronomers, 
enumerated almost a decade ago \citep{2015aska.confE..36K}, 
has also expanded. Moreover, practical 
considerations led to a revised design of SKAO telescopes since 2015, with  
a nominal loss of 50 \% and 30\% in collecting area for the SKA-Low 
and the SKA-Mid telescopes respectively. With the construction 
of the re-baselined telescopes well underway, the pulsar community 
can better define the expected science outcomes from these instruments. 
A brief overview  of the evolving landscape of pulsar astronomy enabled by 
the upcoming SKAO telescopes is presented 
in this article introducing the main themes reviewed 
in subsequent articles of this special issue.


\section{The SKAO telescopes and pulsar observations}

The SKAO telescopes are conceived as two telescopes geographically stationed 
on different continents and with different designs. The SKA-Low telescope 
is located in Australia, whereas the SKA-Mid is coming up in South Africa. Together, the two 
telescopes cover an observing frequency range of 50 MHz to 
15 GHz. The two telescopes will be commissioned and operated by 
an inter-governmental organisation bringing together more than 13 nations 
from around the world in a unique collaborative international effort. 

The construction plan of the SKAO telescopes is envisaged as a staged 
delivery of evolving array assembly (AA). The earliest milestone, AA*, 
is likely be delivered between 2029 to 2031, and the design 
baseline AA4 will follow depending on 
funding\footnote{Please refer to \href{https://www.skao.int/en/science-users/599/scientific-timeline}
{https://www.skao.int/en/science-users/599/scientific-timeline} for details.}. 
We briefly summarise the main features of the two SKAO telescopes below 
highlighting the capabilities important for pulsar science:

{\it SKA-Low:} The telescope site is in Western Australia's Murchison 
shire on the traditional lands of the Wajarri Yamaji and has been 
designed to operate at low radio frequencies, between $50-350\,$MHz. 
Unlike typical dishes, the SKA-Low telescope is built from `wired' 
log-periodic antennas, a design which is efficient at low radio 
frequencies. Each SKA-Low station consists of  256 such antennas equivalent 
to a 39-m single dish with an expected primary beam of 2.2 degrees at 200 MHz. 
The AA* configuration is planned to have 307 stations, rising to 512 stations for AA4, 
with a maximum separation of $74\,$ km between antenna stations.  
The number of closely spaced stations within a 530-m core of 
the interferometer is 224. The rest will be arranged in a spiral arm 
geometry, extending outwards from the core. The AA* will have  199 core 
stations, whereas AA4 will have the full complement of 224 core stations.  
From the point of view of pulsar observations, the output of these stations 
can be combined in phase as a tied array, where independently 
steerable multiple dual polarization beams can be formed within the primary beam 
of each station as explained later. The availability of multiple beams within 
the wide primary beam has a significant impact on pulsar science case as would 
be illustrated in the accompanying articles of this special issue.

{\it SKA-Mid:} This telescope is located in the Karoo region of South Africa. 
The array will use offset Gregorian dishes to cover an observing band 
from $0.35-15.4\,$GHz with a target extension of up to $24\,$GHz. Thus, 
the SKA-Mid telescope's frequency range complements the SKA-Low observing 
band. The telescope will subsume the already commissioned dishes of 
the MeerKAT telescope, which has been operating as a pathfinder 
telescope for the SKAO for the last 6 years at the same site. In the AA* 
configuration, the array will consist of 144 dishes, comprising 64 
existing 13.5-meter MeerKAT dishes and 80 new 15-meter SKA dishes. 
In the AA4 configuration, the full array will include 197 dishes, with 
the number of SKA dishes increasing from 80 to 133. The spatial distribution 
of the dishes is similar to the SKA-Low telescope, where $\sim$ 50\% of the 
dishes will form the central core of $\sim$2 km in extent, while the remaining 
50\% will emerge out of the central core in three spiral arms spanning a 
maximum baseline of $150\,$km. Given the higher observing frequencies 
available in the SKA-Mid telescope, the primary beam is much smaller than that in the SKA-Low and progressively reduces for higher 
frequency bands. The observing frequencies from $0.35-15.4\,$GHz are 
covered with 5 bands (Band 1 to 5) with bandwidths ranging from  
700 to 2500 MHz. The most relevant bands for the pulsar science case 
are Band 1 (350$-$1050 MHz) and Band 2 (950$-$1760 MHz), although 
Band 4 and 5 will be relevant for the science case related 
to the Galactic centre survey and follow up observations \citep{abbate2025}. 
Similar to the SKA-Low, the output of these stations can be combined in phase 
as a tied array, where independently steerable multiple beams can be 
formed within the primary beam. 

The SKAO is an interferometer well suited for imaging observations. A small 
part of the pulsar science case does involve such imaging observations \citep{gelfand2025}, 
but the major fraction of pulsar science will involve non-imaging data 
processing requiring specialised hardware and software for processing 
raw data spectral time-series. The SKAO will utilise two non-imaging 
data processing components for pulsar science, which are briefly described below. 
 
\emph{The Pulsar Timing Subsystem (PST)} is part of SKAO's Central Signal Processor 
(CSP) and is responsible for the high-precision timing and sensitive observations 
of pulsars. During AA*, the PST will support the processing of up to 8 and 16 
dual-polarisation wide-band tied-array beams for the SKA-Low and the SKA-Mid 
telescopes, respectively, using input from the correlator-beamformer (CBF). 
By AA4, both the SKA-Low and the SKA-Mid telescopes are expected to handle 
up to 16 simultaneous beams. The corresponding bandwidths will be $300\,$MHz 
for the SKA-low and up to $2500\,$MHz for the SKA-Mid telescope. Depending 
on the configuration, these tied array beams can be distributed across 
up to 16 subarrays. In order to support a range of scientific use cases,  
the PST will be capable of generating a suite of 
different data products, including channelised dual-polarisation voltage 
time-series, time and frequency resolved PSRFITS filterbanks, and 
integrated pulse profiles.

\emph{The Pulsar Search Subsystem (PSS)} is another key component 
of SKAO's CSP. It is designed to carry out near real-time searches 
of the sky visible from SKAO's sites for pulsars and transient 
radio sources. The PSS combines custom hardware and dedicated software 
optimised for high-throughput, real-time signal processing. During AA*, 
it will support the parallel processing of up to 250 and 1125 tied-array 
beams for the SKA-Low and the SKA-Mid telescopes, respectively, 
increasing to 500 and 1500 beams by AA4. The processed bandwidths 
will be $100\,$MHz for the SKA-Low and up to $300\,$MHz for the SKA-Mid 
telescope. The SKA-Mid beamformer will supply channelised, full-polarisation 
data to the PSS, while the SKA-Low beamformer will provide 
dual-polarisation voltage data. Beams may be flexibly grouped 
into up to 16 independent subarrays, subject to some restrictions. The PSS will perform pulsar 
searches on integrations of up to 1800 s and frequency-domain acceleration 
searches in integration times of up to 10 minutes. In parallel, 
the single-pulse search pipeline will identify astrophysical 
transient events, operating either alongside the acceleration search 
or commensally during non-pulsar observations.

In a recent development, some of the above mentioned capabilities of the 
SKAO telescopes have been deferred for a later date. These include 50 
fewer stations in the SKA-Low telescope core for AA* and reduction in the 
number of beams of the PSS. These capabilities will be restored at a later 
date. While this does not impact the overall science outcome for pulsar 
science described in this special issue, it may impact the time-line of 
achieving these goals in a major way. Therefore, it is highly desirable 
to reinstate these capabilities as early as possible in the roll out 
of the SKAO telescopes.

The SKAO telescopes will provide unprecedented sensitivity over 50 MHz to 15 GHz. 
The SKA-Mid telescope will be around three to four times
more sensitive than MeerKAT in the AA* and AA4 configurations, respectively. 
A similar progression is expected for the SKA-Low telescope, which will 
eventually surpass the sensitivity of the low-frequency array LOFAR 
by nearly an order of magnitude once fully deployed. The ability to 
flexibly combine the stations and the antennas in the 
SKA-Low and SKA-Mid telescopes as multiple subarrays assigned to different 
beams provided by the PST, coupled with large number of beams in the PSS within 
the relatively large primary beam of the telescopes, along with near real-time search 
capability and capability to produce a variety of reduced data products, will be a 
game-changer for a variety of pulsar science when 
combined with an order of magnitude increase in instantaneous sensitivity 
of SKAO telescopes over the current instruments. 


\section{State of the art and expected SKAO discoveries}

\begin{figure*}[t] 
  \centering
  \includegraphics[scale=0.5]{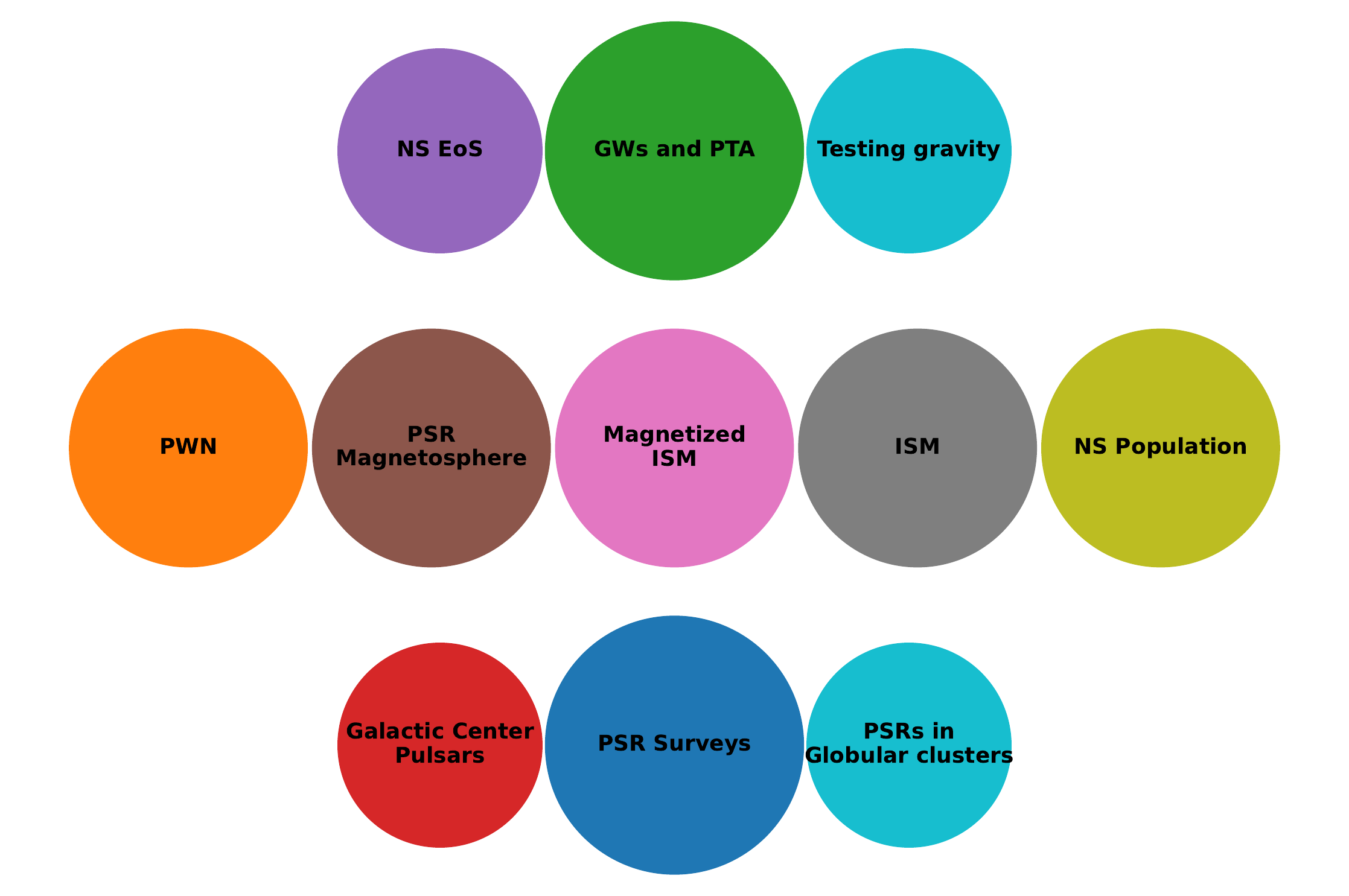}
  \caption{The inter-connected pulsar science use cases for the SKAO, described in this 
  special issue, are depicted here. The surveys shown at the bottom form 
  the backbone of the study of different areas depicted in the middle row. 
  The enhanced understanding in the latter will feed into the three main 
  contributions to fundamental physics from pulsar astronomy, which are shown in the top row.}
  \label{fig:psrscience}
\end{figure*}

There have been significant advances in pulsar astronomy since the pulsar science 
case for the SKAO was drafted two decades ago, particularly since its update a 
decade back \citep{2015aska.confE..36K}. The gravitational waves at high 
frequencies have been detected from more than 200 sources and the pulsar timing 
arrays are close to a significant detection of nano-Hertz gravitational waves 
opening up a messenger that the SKAO telescope can explore in earnest. 
More exotic binary systems have been discovered and mass and radius for 
several NS are measured to a good precision motivating investigations in 
fundamental physics using pulsars as probes. The large instantaneous 
sensitivity and flexibility of the SKAO telescope will make a significant 
impact on the pulsar science case with doubling the pulsar population 
in a pulsar survey, which forms the basis for fundamental physics 
studies interconnected with other areas of pulsar astronomy as illustrated 
in Figure \ref{fig:psrscience}. A brief overview of the current status of 
these science cases with anticipated impact of the SKAO is presented in the 
next few sections with the details elaborated 
in the accompanying articles in this special issue.

\subsection{The SKA Pulsar Census (Keane et al.)}

Over the past fifty years, pulsar surveys have expanded the known population to over 4000 sources, revealing extraordinary diversity in spin, magnetic field, and evolutionary stage. Yet, population synthesis models predict that the Milky Way hosts more than 50,000 active pulsars. Even with modern facilities such as Murriyang (the Parkes radio telescope), FAST, and MeerKAT, much of this population---particularly faint, distant, or fast-spinning pulsars---remains undetected. A complete census of pulsars is the foundation for nearly all SKA pulsar science, underpinning studies of Galactic structure, neutron-star physics, and gravitational-wave detection.

The \textbf{SKA-Low AA*} and \textbf{SKA-Mid AA*} arrays will form a coordinated two-tier survey strategy. The SKA-Low telescope will sweep the Galactic halo and high-latitude regions at low frequencies, while the SKA-Mid telescope will target the dense Galactic plane where dispersion and scattering are strongest. Together, they are expected to detect approximately 10,000  slow pulsars and nearly 800 millisecond pulsars (MSPs), more than doubling the known population. This early delivery phase will already establish a statistically robust sample for pulsar timing array and population studies.

The full \textbf{AA4} configuration will increase sensitivity by 20–30\%, raising the total yield above 12,000 pulsars and providing complete coverage of the Galactic disk, bulge, and halo. The \textbf{SKA-Mid AA4} will deliver ultra-precise timing of high-dispersion and relativistic systems, while the  \textbf{SKA-Low AA4} will identify older, faint, and diffuse sources in the outer Galaxy. This comprehensive census will serve as the foundation for all subsequent SKA pulsar investigations, forming the most complete neutron-star catalogue in existence \citep{keane2025}.

\subsection{Pulsars in Globular Clusters (Bagchi et al.)}

Globular clusters (GCs) are stellar systems with extreme densities, where frequent stellar encounters lead to efficient formation of millisecond pulsars through binary recycling. Surveys with Arecibo, Parkes, and FAST have discovered 345 cluster pulsars, some probing exotic regimes such as pulsar--black-hole binaries and ultradense stellar cores. Yet, modelling suggests that thousands of cluster MSPs remain hidden, limited primarily by telescope sensitivity.

The \textbf{SKA-Mid AA*} array will make immediate breakthroughs through targeted, deep searches of Galactic clusters, where only a few tied-array beams are required. Early observations are expected to more than double the number of known cluster pulsars by detecting faint and distant MSPs previously below the detection threshold. The \textbf{SKA-Low AA*} will complement these searches at lower frequencies, identifying steep-spectrum sources and diffuse emission halos around cluster cores.

Once the full \textbf{AA4} arrays are operational, the \textbf{SKA-Mid AA4} could increase the total number of detected cluster pulsars by a factor of four to five---up to 1,700 pulsars across the Galactic GC population. The \textbf{SKA-Low AA4} will simultaneously monitor multiple clusters to measure dispersion and rotation measures, mapping intra-cluster gas and magnetic fields. These data will enable precision timing of exotic binaries, exploration of cluster evolution, and tests of gravity in the densest stellar environments \citep{bagchi2025}.

\subsection{Galactic Centre Pulsars (Abbate et al.)}

The Galactic Centre, hosting Sagittarius~A*, is one of the most extreme gravitational environments known. Infrared monitoring of orbiting stars has confirmed the black hole’s mass and Schwarzschild precession, but pulsars---ideal clocks in curved spacetime---have yet to be found in close orbit. Despite decades of surveys with Parkes, Effelsberg, and the Green Bank Telescope, only seven pulsars within 100~pc of the Centre are known, their detection hindered by severe radio scattering.

The \textbf{SKA-Mid AA*} configuration will overcome this barrier through its high-frequency coverage (up to 15~GHz) and sensitivity, capable of detecting heavily scattered pulsar signals near Sgr~A*. Meanwhile, the \textbf{SKA-Low AA*} will map the intervening ionized medium and measure scattering properties. 
Even during early operations, AA* is expected to uncover dozens of new pulsars in the region, expanding the sample dramatically.

At full power, the \textbf{SKA-Mid AA4} will time pulsars orbiting Sgr~A* with microsecond precision, measuring its spin and quadrupole moment---yielding direct tests of the cosmic censorship and no-hair theorems. The \textbf{SKA-Low AA4} will expand the survey area to identify 
bright transients/pulsars in the surrounding star-forming region. Together, these discoveries will transform our understanding of the Galactic Centre’s plasma physics, gravitational field, and possibly its dark-matter content \citep{abbate2025}.

\subsection{The Galactic Neutron Star Population (Levin et al.)}

The past half century has revealed that neutron stars manifest in many forms---ordinary pulsars, magnetars, rotating radio transients (RRATs), intermittent pulsars, central compact objects (CCOs), and millisecond pulsars (MSPs), many of the latter in exotic binary systems, such as double neutron stars, 
 triple system and transitional pulsars---each providing insight into extreme physics. However, biases toward bright, long-lived sources have obscured the true diversity of this population. Many faint, intermittent, or short-lived neutron stars remain undetected, preventing a unified view of their life cycles.

The \textbf{SKA-Low AA*} and the \textbf{SKA-Mid AA*} telescopes will bridge these gaps by discovering and characterizing thousands of new neutron stars across all subgroups. The \textbf{SKA-Low AA*}, with its broad field of view, will be ideal for detecting faint and steep-spectrum sources, while the  \textbf{SKA-Mid AA*} will enable high-cadence timing and polarization studies to classify and connect subpopulations. Together, these facilities will reveal transitional objects---linking, for example, high-B pulsars to magnetars or RRATs to ordinary pulsars---and provide the first coherent picture of neutron-star evolution.

In the \textbf{AA4} era, \textbf{SKA-Mid} will conduct continuous timing and polarization monitoring of thousands of sources, measuring glitches, emission variability, and magnetospheric changes, while \textbf{SKA-Low} will expand discovery surveys into the Galactic halo. These combined efforts will yield the first statistically complete neutron-star census, tracing evolutionary pathways and unifying disparate classes within a single population framework \citep{levin2025}.

\subsection{Pulsar Magnetospheres (Oswald et al.)}

Advances in radio and high-energy astrophysics have revealed that pulsar magnetospheres exhibit extraordinary complexity, including non-dipolar fields, plasma instabilities, and emission variability. Observations with FAST, MeerKAT, and NICER have deepened our understanding but left the core questions---how and where coherent radio emission arises---unanswered.

The \textbf{SKA-Mid AA*} array, with its wide bandwidth and unmatched frequency coverage,
particularly using sub-arrays for concurrent multi-band observations together with its  large sensitivity, will conduct high-cadence monitoring of polarization and single-pulse behavior in hundreds of pulsars. This will constrain magnetic geometry and emission altitude, while the \textbf{SKA-Low AA*} will probe coherent plasma processes near the stellar surface. Together, the SKA-Low and SKA-Mid arrays will bridge observational and theoretical gaps in pulsar emission physics.

In the \textbf{AA4} configuration, the \textbf{SKA-Mid} will monitor thousands of pulsars at sub-millisecond resolution, uncovering population-wide variability patterns, while the \textbf{SKA-Low} will capture weak, intermittent, or low-frequency emission signatures. These results will refine models of magnetospheric structure, test theories of particle acceleration, and link observed radio emission directly to fundamental plasma dynamics \citep{oswald2025}.

\subsection{Galactic Plasma Studies (Tiburzi et al.)}

Pulsars have long served as precise probes of the Galaxy’s ionized plasma. Observations with LOFAR, GMRT, and MeerKAT have traced electron-density fluctuations and revealed turbulence consistent with a Kolmogorov spectrum, while long-term DM monitoring has exposed changes due to solar wind and interstellar structures. Yet, these measurements are limited in frequency coverage and spatial sampling.

The \textbf{SKA-Low AA*} array will achieve unmatched precision in dispersion and scattering studies, tracking DM variations across hundreds of channels and enabling detailed mapping of the turbulent interstellar medium. In parallel, the \textbf{SKA-Mid AA*} will provide high-frequency coverage to separate frequency-dependent dispersion from refraction, allowing a full multi-scale characterization of Galactic plasma and solar-wind dynamics.

Once \textbf{AA4} arrays are in place, the \textbf{SKA-Low} will deliver a high-resolution 3D model of Galactic electron density, while the \textbf{SKA-Mid} will furnish ultra-stable timing data critical to Pulsar Timing Arrays. Together, they will refine models of plasma turbulence, correct for ISM effects in gravitational-wave searches, and provide the most comprehensive view to date of the Milky Way’s magneto-ionic medium \citep{tiburzi2025}.

\subsection{Galactic Magnetic Fields (Xu et al.)}

The study of Galactic magnetism has progressed from optical polarization measurements to radio Faraday rotation mapping using pulsars and extragalactic sources. Large scale experiments with FAST, MeerKAT and Parkes have revealed large-scale field reversals and magnetized halo structures, yet data remain confined to the local half of the disk. The three-dimensional structure and global symmetry of the Milky Way’s magnetic field remain incompletely characterized.

The \textbf{SKA-Mid AA*} and \textbf{SKA-Low AA*} arrays will transform this picture. Early-phase (AA*) observations will triple the number of pulsars with measured rotation measures, providing thousands of new magnetic-field sight-lines across the disk and halo. The \textbf{SKA-Mid AA*} will target the inner spiral arms, while the \textbf{SKA-Low AA*} will probe the diffuse magnetism of the Galactic halo and high-latitude sky, extending magnetic mapping to kiloparsec scales.

With the full \textbf{AA4} arrays, the SKAO telescopes will deliver a much improved 3D magnetic-field map of the Milky Way. The \textbf{SKA-Mid AA4} will measure field reversals and small-scale turbulence, and the \textbf{SKA-Low AA4} will chart the extended halo field.


\subsection{Pulsar Wind Nebulae (Gelfand et al.)}

Pulsar wind nebulae (PWNe) illustrate how neutron stars transfer rotational energy to their surroundings, producing non-thermal emission across the electromagnetic spectrum. Observations with Chandra, H.E.S.S., and MeerKAT have revealed spectacular structures such as jets, tori, and bow shocks, but most PWNe remain unresolved, and their particle acceleration mechanisms are not yet fully understood.

The \textbf{SKA-Mid AA*} array will provide deep, wide-band imaging and polarization mapping of PWNe, tracing their magnetic topology and energy distribution. \textbf{SKA-Low AA*} will detect diffuse, low-frequency synchrotron emission from older or larger nebulae, capturing relic populations of relativistic electrons. Joint SKA-Low and SKA-Mid observations will reveal how magnetic structure and particle spectra evolve over time.

At full capacity, \textbf{SKA-Mid AA4} will resolve fine-scale features---filaments, shocks, and jets---at sub-arcsecond resolution, while \textbf{SKA-Low AA4} will quantify large-scale halos and measure depolarization to infer field geometry. Together, these data will clarify how pulsar winds accelerate particles to PeV energies, their role in cosmic-ray production, and their feedback on the interstellar medium \citep{gelfand2025}.

\subsection{Testing Gravity with Binary Pulsars (Venkatraman Krishnan et al.)}

Binary pulsars are unique laboratories for testing general relativity (GR) and alternative theories of gravity in the strong-field regime. From the orbital decay of the Hulse--Taylor pulsar to the double pulsar PSR~J0737--3039A/B, such systems have validated gravitational-wave emission and post-Keplerian dynamics to extraordinary precision. However, only a limited number of relativistic binaries are known, and pulsar--black-hole systems---crucial for probing spacetime curvature and the no-hair theorem---have yet to be discovered.

The \textbf{SKA-Mid AA*} array will revolutionize this field by increasing timing precision by an order of magnitude and detecting many new relativistic binaries. Early AA* observations will measure minute effects such as Shapiro delay, gravitational redshift, and frame dragging in known systems. In parallel, the \textbf{SKA-Low AA*} will survey for long-period and eccentric binaries, populating the full parameter space of relativistic systems for gravity tests. The timing precision will also be significantly improved incorporating systematics due to propagation effects in the inter-stellar 
medium using the lower frequencies provided by both the \textbf{SKA-Mid} and \textbf{SKA-Low} telescopes. 

The large instantaneous sensitivity of \textbf{AA*} and \textbf{AA4} telescopes will lead to 
discovery of systems with orbital period shorter than 2 hours and higher eccentricities,
probing larger curvature parameter space in relatively shorter observing campaigns. They will also allow for more than
a factor of two improvement in ToA uncertainty and an order of magnitude improvement in the precision of 
Post-Keplerian (PK) parameters, even with integration time as small as 30 s. With the \textbf{AA4} upgrade, \textbf{SKA-Mid} will achieve tens-of-nanosecond timing accuracy, allowing detection of higher-order post-Newtonian corrections and direct verification of GR’s predictions for black-hole multipole moments. \textbf{SKA-Low AA4} will extend the search to the Galactic periphery, identifying rare systems that complement timing-array experiments. Combined, these datasets will establish SKA as the premier facility for experimental gravitation, exploring the limits of Einstein’s theory with unmatched precision \citep{venkatraman2025}.

\subsection{Probing Neutron Star Interiors and Dense Matter (Basu et al.)}

Observations over the last decade have established neutron stars as laboratories for ultra-dense matter physics at relatively low temperatures and large proton-neutron asymmetries. NICER and gravitational-wave detections along with radio timing observations of binary neutron stars have constrained masses, radii and moment of inertia. The timing studies of young pulsars have revealed superfluidity, crustal entrainment  and internal coupling. Nevertheless, the microphysics of matter above nuclear saturation density---and the transition between hadronic and exotic phases---remains uncertain.

The \textbf{SKA-Mid AA*} system will immediately improve mass and spin measurements by timing massive and rapidly rotating pulsars at sub-microsecond precision. The \textbf{SKA-Low AA*} will track glitches and precession events along with variation in propagation delays in the inter-stellar medium, providing higher precision measurements of these quantities. It will also allow us to better characterize the timing noise of these young pulsars, thererby revealing superfluid dynamics within neutron-star interiors. Early AA* results will narrow the range of viable equations of state (EoS) and test theoretical models of nuclear interactions and phase transitions.

At full \textbf{AA4} sensitivity, the \textbf{SKA-Mid} will enable high precision moment-of-inertia measurements in relativistic binaries and explore frame-dragging effects linked to internal structure. The \textbf{SKA-Low AA4}, operating at lower frequencies, will characterize glitch recovery and long-term timing noise, probing superfluid vortices and crust elasticity. An observing program resolving rotational variations on timescales from seconds to years with dynamic high cadence scheduling 
near a glitch for a large sample of pulsars with \textbf{AA4} is planned and commensality with other observations is recommended. The mass priors and geometric/polar cap priors as well as precision radio ephmerides provided by \textbf{AA4} telescopes will benefit 
multi-messenger observations with upcoming facilities, such as eXTP, New Athena, LISA, Cosmic 
explorer and Einstein telescope to provide tight constraints on the EoS. Therefore, 
the SKAO telescopes will provide definitive observational constraints on the EoS and the behavior of matter at supra-nuclear densities \citep{basu2025}.

\subsection{The SKAO Pulsar Timing Array (Shannon et al.)}

The spatial correlation introduced by gravitational waves (GWs) in the precision timing of an ensemble of millisecond pulsars, or a pulsar timing array (PTA), provides a complementary view of universe to that seen in decahertz GWs by terrestrial detectors, such as advanced LIGO. While a compelling evidence was evident in decade long data released by the existing PTA experiments two years back, a significant detection -- and indeed understanding the origins of these GWs -- is yet to happen. In future, PTAs 
are likely to be able to discriminate between an astrophysical or cosmological origin of the stochastic gravitational background, validate and refine hierarchical galaxy evolution models, confirm or rule out anisotropy in the GW  background, set stringent limits on the dipolar GW radiation and discriminate between alternative gravity theories with a rich dividend for both astrophysics and fundamental physics. This requires extending the current time baseline in the coming decades, as well as increasing the sensitivity of a PTA experiment. This was the prime aim of pulsar science case for the SKAO telescopes, bringing together gains from  sensitive pulsar surveys and advances in pulsar EoS, inter-stellar medium and pulsar emission physics.

The \textbf{SKA-Mid} and \textbf{SKA-Low} telescopes would play a pivotal role in this respect. The \textbf{SKA-Mid AA*} array would increase the MSP ensemble by either extending the time baselines for the currently monitored MSPs at the MeerKAT, Parkes pulsar timing array, North American Nano-Hertz Gravitational wave Observatory, European pulsar timing array, the Indo-Japanese pulsar timing array and the Chinese pulsar timing array or by including hitherto inaccessible weaker MSPs. The sample will be augmented by new discoveries in the pulsar census \citep{keane2025} with the \textbf{SKA-Mid AA*} and \textbf{SKA-Low AA*}. The SKAO PTA will also benefit by better modelling of chromatic effects using wide frequency coverage provided by both the \textbf{SKA-Mid} and \textbf{SKA-Low} telescopes apart from better characterization of pulse jitter. 

With the full \textbf{AA4}, the \textbf{SKA-Mid} and \textbf{SKA-Low} telescopes will further provide  sub-$\mu$s precision times-of-arrival for more than double the number of currently monitored pulsars with a modest use of SKAO telescope time, largely due to a factor of four times the MeerKAT sensitivity. 
The lower observing frequency band of the \textbf{SKA-Low AA4} array will better characterize the chromatic delays, pulse-shape evolution, and DM variability thereby accounting for the noise budget more accurately. More accurate distance measurements with VLBI measurements using the SKAO telescopes will help in better localization of individual continuous GW sources \citep{shannon2025}.

\section*{Summary}

Together, these studies define the pulsar and neutron-star science programme for the SKA Observatory. The \textbf{AA*} arrays will provide an immediate leap in discovery power and precision, while the full \textbf{AA4} configurations of \textbf{SKA-Low} and \textbf{SKA-Mid} will deliver transformative, population-level insights into neutron-star physics, plasma astrophysics, Galactic structure, and gravitation.


\section*{Acknowledgements}

The authors acknowledge useful discussion and inputs from Dr Maciej Serylak, Dr Willem van Straten and Dr Lina Levin Preston about the SKAO telescopes and their sub-systems. BCJ acknowledges the support from Raja Ramanna Chair fellowship of the Department of Atomic Energy, Government of India (RRC - Track I Grant 3/3401 Atomic Energy Research 00 004 Research and Development 27 02 31 1002//2/2023/RRC/R\&D-II/13886 and 1002/2/2023/RRC/R\&D-II/14369). MB acknowledges resources
from the research grant “iPeska” (PI: Possenti), funded under the INAF national call Prin-SKA/CTA approved with the Presidential Decree 70/2016.

\bibliographystyle{apalike}
\bibliography{ska_pulsars}
\end{document}